\documentclass[conference,10pt]{IEEEtran}
\IEEEoverridecommandlockouts
\usepackage[utf8]{inputenc} 
\usepackage[T1]{fontenc}
\usepackage{url}              
\usepackage{cite}             

\usepackage[cmex10]{amsmath} 
\usepackage{mleftright}      
\mleftright                  
\usepackage{amsmath}
\DeclareMathOperator*{\argmin}{argmin}
\usepackage{graphicx}         
\usepackage{booktabs}         
\usepackage{mathtools}
     
\usepackage{graphicx,cite,amsmath,amssymb,amsfonts,mathtools,algorithm,mathabx,textcomp,blindtext,xcolor,subfig,float,mwe,dblfloatfix,bm, xfrac,algpseudocode, stackengine, verbatim,xparse} 
\usepackage{amsthm}

\newtheorem{proposition}{Proposition}

\def\BibTeX{{\rm B\kern-.05em{\sc i\kern-.025em b}\kern-.08em
    T\kern-.1667em\lower.7ex\hbox{E}\kern-.125emX}}
\begin{document}
\title{Coverage Analysis of Multi-Environment Q-Learning Algorithms for Wireless Network Optimization\\
\thanks{This work was funded by the following grants: NSF CCF-1817200, ARO W911NF1910269, DOE DE-SC0021417,
Swedish Research Council 2018-04359, NSF CCF-2008927, NSF CCF-2200221, ONR 503400-78050, ONR N00014-15-1-2550, NSF CCF 2200221 and USC + Amazon Center on Secure and Trusted Machine Learning}
}
\IEEEaftertitletext{\vspace{-3\baselineskip}}
\author{\IEEEauthorblockN{Talha Bozkus}
\IEEEauthorblockA{\textit{Department of Electrical and Computer Engineering} \\
\textit{University of Southern California, USA}\\
bozkus@usc.edu}
\and
\IEEEauthorblockN{Urbashi Mitra}
\IEEEauthorblockA{\textit{Department of Electrical and Computer Engineering} \\
\textit{University of Southern California, USA} \\
ubli@usc.edu}
}

\maketitle
\begin{abstract}
Q-learning is widely used to optimize wireless networks with unknown system dynamics. Recent advancements include ensemble multi-environment hybrid Q-learning algorithms, which utilize multiple Q-learning algorithms across structurally related but distinct Markovian environments and outperform existing Q-learning algorithms in terms of accuracy and complexity in large-scale wireless networks. We herein conduct a comprehensive coverage analysis to ensure optimal data coverage conditions for these algorithms. Initially, we establish upper bounds on the expectation and variance of different coverage coefficients. Leveraging these bounds, we present an algorithm for efficient initialization of these algorithms. We test our algorithm on two distinct real-world wireless networks. Numerical simulations show that our algorithm can achieve \%50 less policy error and \%40 less runtime complexity than state-of-the-art reinforcement learning algorithms. Furthermore, our algorithm exhibits robustness to changes in network settings and parameters. We also numerically validate our theoretical results.
\end{abstract}

\setlength{\textfloatsep}{1pt}
\section{Introduction}\label{sec:introduction}

Markov Decision Processes (MDPs) are useful tools for modeling large-scale wireless networks and solving diverse wireless network optimization problems such as throughput maximization, collision avoidance, and route optimization \cite{talha_jie_asilomar, colink_journal, ln_journal, talha_asilomar}. When transition dynamics and cost functions are unknown, traditional methods such as dynamic programming \cite{bertsekas_book} are inapplicable. Instead, \textit{model-free} reinforcement learning (RL) algorithms such as $Q$-learning \cite{bertsekas_book} can be employed. However, traditional $Q$-learning faces several challenges, including high computation complexity, high estimation bias and variance, and slow convergence. Consequently, several variants have been proposed to address these issues. These algorithms can be classified based on their estimation strategies, objectives, and implementations. 

Similar to the original Q-learning, a single Q-function estimator on a single Markovian environment is employed in \cite{speedy_q, fitted_q, bertsekas_book}. On the other hand, multiple Q-function estimators on a single Markovian environment are used in \cite{double_q, averaged_dqn, ensemble_bootstrap_q, randomized_double_q}, in which each estimator is initialized independently, and their outputs are fused into a single estimate via some weighting mechanism. Recently, novel ensemble Q-learning algorithms that utilize multiple Q-function estimators on multiple distinct but structurally related Markovian environments have been proposed \cite{pn_journal, ln_journal}. These algorithms run multiple Q-learning algorithms on multiple Markovian environments. These environments are inherently different but share similar characteristics and dynamics, enabling improved performance with respect to exploration and training \cite{pn_journal}, \cite{ln_journal}. The work of \cite{pn_journal} addresses MDPs with arbitrary graphs and presents probabilistic convergence analysis by assuming distributional assumptions on Q-functions. In contrast, \cite{ln_journal} leverages MDP structural properties to reduce training complexity with efficient sampling and state-aggregation strategies and provides deterministic stability and convergence analysis. Q-learning algorithms can also be categorized based on their objective. Estimation bias is considered in \cite{double_q}, and the estimation variance and training stability are examined in \cite{averaged_dqn, randomized_double_q}. The convergence rate is improved in \cite{speedy_q}, and training data efficiency is considered in \cite{fitted_q}. The work of \cite{pn_journal, ln_journal} improves the overall exploration across large state-spaces and reduces the training runtime and sample complexity -- the number of interactions with the environment to learn optimal policies.

$Q$-learning algorithms also differ in terms of their implementation and data collection strategies. In online Q-learning \cite{online_q_learning_1, online_q_learning_2}, the RL agent learns and updates its Q-functions by interacting with the environment in real-time. This approach is suitable for dynamic environments. In contrast, offline Q-learning \cite{offline_q_learning_1, offline_q_learning_2} relies on pre-collected training data and is preferred when direct interaction with the environment is expensive. Recent hybrid versions \cite{hybrid_q_1, hybrid_q_2, pn_journal, ln_journal} combine online interaction with offline data for improved performance. 

Despite their distinct characteristics, reducing sample complexity remains a crucial consideration across different types of Q-learning algorithms. To this end, the notion of \textit{coverage conditions} plays a fundamental role in determining the sample complexity of RL algorithms \cite{coverage_1, coverage_2}. These conditions ensure that the data collection distribution adequately covers the state space. For instance, Fitted Q-Iteration \cite{fitted_q} attains an $\epsilon$-optimal policy under the condition that the data distribution uniformly covers all possible induced state distributions.

In this work, we will analyze the coverage conditions of recent multi-environment Q-learning algorithms \cite{pn_journal, ln_journal}, which have not been well-studied. We aim to improve the accuracy and complexity of these algorithms by leveraging their theoretical coverage limits. Our motivation is to handle complex scenarios in optimizing real-world wireless networks, such as multiple packet arrivals to energy-harvesting transmitters, significant channel quality variations in MIMO networks, simultaneous packet collisions, or buffer overflows due to multiple data arrivals in MISO networks.

The main contributions of the paper are as follows: (i) We adopt a probabilistic approach to derive upper bounds on the expectation and variance of different coverage coefficients (CC) for multi-environment ensemble Q-learning algorithms. To the best of our knowledge, there is no prior work on CC that takes this approach. (ii) We present an algorithm to assess and compare the utilities of different Markovian environments within these algorithms. Unlike prior work \cite{pn_journal}, our approach offers more accurate, low-complexity, and robust initialization for these algorithms. (iii) We conduct simulations on two real-world large state-space wireless networks, showing that our algorithm reduces policy error by 50\% and increases speed by 40\% compared to state-of-the-art RL algorithms. We also validate our theoretical findings numerically. 

In this paper,  vectors are bold lower case (\textbf{x}), matrices are bold upper case (\textbf{A}), and sets are in calligraphic font ({$\mathcal{S}$}).

\section{System Model and Tools}
\label{sec:system_model}

\subsection{Markov Decision Processes}
\label{subsec:mdp}
MDPs are characterized by 4-tuples $\{\mathcal{S}$, $\mathcal{A}$, $p$, $c$\}, where $\mathcal{S}$ and $\mathcal{A}$ denote the finite state and action spaces, respectively. We denote $s_{t}$ as the {\em state} and $a_{t}$ as the {\em action} taken at discrete time $t$. The transition from $s$ to $s^{\prime}$ under action $a$ occurs with the probability $p_{a}(s,s^{\prime})$, which is stored in the $(s,s',a)^{th}$ element of the probability transition tensor (PTT) $\mathbf{P}$, and a bounded cost $c_{a}(s)$ is incurred, which is stored in the $s^{th}$ element of the cost vector $\mathbf{c}_a$. We focus on an infinite-horizon discounted cost MDP, where $t = \mathbb{Z}^{+}\cup \{0\}$. Our goal is to solve the following optimization problem: 
\begin{equation}\label{Equ: optimization_eq}
\mathbf{v}^{*}(s)=\min _{\pi} \mathbf{v}_{\pi}(s)=\min _{\pi}\mathbb{E}_{\pi}\Big[\sum_{t=0}^{\infty} \gamma^{t} \mathbf{c}_{a_{t}}(s_{t}) | s_{0}=s\Big],
\end{equation}
for all $s \in \mathcal{S}, a_{t} \in \mathcal{A}$, where $\mathbf{v}_{\pi}$ is the \textit{value function} \cite{bertsekas_book} -- the expected cumulative discounted cost incurred starting from state $s_0$ under the \textit{policy} ${\pi}$, $\mathbf{v}^{*}$ is the \textit{optimal value function}, and $\gamma \in (0,1) $ is the discount factor. The \textit{deterministic} policy defines a specific action per state, and the \textit{stationary} policy does not change over time. We herein consider deterministic and stationary policies as they always exist given a finite state and action space \cite{bertsekas_book}.

\subsection{Q-Learning}\label{subsec:q_learning}
Q-learning seeks to solve the optimization problem and find the optimal policy $\pi^{*}$ by learning the Q functions $Q(s,a)$ -- the expected cumulative discounted cost of taking action $a$ in state $s$, $\forall (s,a) \in \mathcal{S} \times \mathcal{A}$ using the following update rule:
\begin{equation}\label{Equ: Q-learning-update-rule}
    Q(s, a) \leftarrow(1-\alpha) Q(s, a)+\alpha(c_{a}(s)+\gamma \min _{a' \in \mathcal{A}} Q(s', a')),
\end{equation}
where $\alpha \in (0,1)$ is the learning rate. The $\epsilon$\textit{-greedy} policies are used to handle the \textit{exploration-exploitation} trade-off: with probability $\epsilon$, a random action is taken (exploration), and with probability 1-$\epsilon$, a greedy action, which minimizes the Q-function of the next state, is chosen (exploitation). This balance is crucial to ensure that sufficient information about the system is captured by visiting each state-action pair sufficiently many times. The agent interacts with the environment and collects samples $\{s,a,s',c\}$ to update Q-functions using (\ref{Equ: Q-learning-update-rule}). The learning strategy must specify the trajectory length ($l$) (the number of states in a single trajectory) and the minimum number of visits to each state-action pair ($v$), which is generally used as a termination condition for the sampling operation. Q-functions converge to their optimal values with probability one, \textit{i.e.,} $Q \xrightarrow{\mathit{w.p.1}} Q^{*}$ if a set of necessary conditions are satisfied \cite{q_learning_convergence}. The policy can then be inferred as $\pi^*(s)=\argmin_{a \in \mathcal{A}} Q^{*}(s, a)$.

\subsection{Coverage Coefficient}\label{coverage_coefficient}

Sample complexity is a crucial consideration for different RL algorithms. In online RL, sample-efficient algorithms require exploration conditions to learn a near-optimal policy when interacting with the unknown environment, such as epsilon-greedy, upper confidence level-based exploration, or Thompson sampling \cite{coverage_3}. In contrast, sample-efficient algorithms for offline RL require data coverage conditions over the offline dataset for statistical guarantees. The work in \cite{coverage_1} shows that the data coverage condition can ensure sample efficiency even in an online setting. The multi-environment Q-learning algorithms \cite{pn_journal, ln_journal} combine features of online and offline RL methods; hence, our focus will be to ensure good data coverage conditions. 

The \textit{coverage coefficient} quantifies data coverage. There exist different definitions for the coverage coefficient (CC) \cite{coverage_2, coverage_3}; we herein use the traditional definition of CC, which is applicable to discrete and finite state-action spaces:\vspace{-2pt}
\begin{align}\label{Equ:CC}
    C^{\pi}(s,a) = \frac{d^{\pi}(s,a)}{v(s,a)}, \quad C^* = \max_{s,a} C^{\pi}(s,a),
\end{align}
where $C^{\pi}(s,a)$ is the CC for the state-action pair $(s,a)$ under the policy $\pi$, $d^{\pi}(s,a)$ is the likelihood of visiting $(s,a)$ when following the policy $\pi$, $v(s,a)$ is the likelihood of visiting $(s,a)$ under some exploratory behavior and $C^*$ is the CC of the overall state-action space. Herein, $C^*$ measures the deviation between the exploration distribution $v$ and the distribution induced by the policy $\pi$. If $v$ is close to $d^\pi$, it indicates that the policy $\pi$ spends a comparable amount of time to the exploration of the state-action pair $(s,a)$ by $v$. Achieving perfect coverage ($C^* = 1$) may not be feasible in practice; however, choosing a good exploration distribution $v$ and optimizing the policy $\pi$ can make $C^*$ close to 1. In this work, we will leverage the distance between the simulated CC from its optimal value across different environments and policies to optimize multi-environment ensemble Q-learning algorithms. 

\section{Theoretical Analysis}
\label{sec:algorithm}

We use Algorithm 2 of \cite{pn_journal} (n-hop ensemble Q-learning = nEQL) as a generic example for our derivations, yet our analysis can be generalized to the algorithm of \cite{ln_journal}.  

\subsection{Assumptions and Preliminaries}\label{Subsec: assumptions}

Let $K$ be the number of environments in the nEQL algorithm, and $n$ denote the order of the environment ($n=1,2,...,K$). We employ the distributional assumption on the $Q$-function errors of the $n^{th}$ environment:
\begin{align}
    \mathcal{X}^{(n)}_t(s,a) = Q_{t}^{(n)}(s,a) - Q^{*}(s,a)  \sim D_n\Big(0,\frac{\lambda_n^2}{3}\Big), \label{Equ: distribution_assumption}
\end{align}
where $\mathcal{X}^{(n)}_t$ is the Q-function error of the $n^{th}$ environment at time $t$, $Q^{*}$ is the optimal Q-function of the original environment, and $Q_{t}^{(n)}$ is the Q-function of the $n^{th}$ environment at time $t$ for $(s,a)$, and $D_n$ is some random distribution (with zero-mean and finite variance). This assumption has been commonly employed in \cite{uniform_assump_1, randomized_double_q} for $n=1$ case with $D_n$ being uniform, non-uniform, or normal. In \cite{pn_journal, ln_journal}, the assumption is generalized to $n>1$ case with no assumptions on $D_n$. We herein present our results for $|\mathcal{A}| = 2$ (see \cite{proof} for the generalization to the arbitrary $|\mathcal{A}|$). We assume that the policy $\pi$ always chooses the action $a$ in (\ref{Equ:CC}). (Otherwise, it is the trivial case: $d^\pi = C^\pi = 0$); thus, $C^\pi \in [1,\infty)$. Furthermore, we employ a linear action selection strategy for each different environment:\vspace{-10pt}
\begin{align}\label{Equ: linear_action_selection}
    v^{(n)}_t(s,a) = \frac{Q^{(n)}_t(s,a)}{\sum_{k=1}^{|\mathcal{A}|} Q^{(n)}_t(s,a_k)},
\end{align}
where $v^{(n)}_t(s,a)$ is the exploration distribution of $(s,a)$ of the $n^{th}$ environment at time $t$. (Softmax-action selection can also be employed, although the analysis is more involved \cite{proof}) We also assume the following relationship between the Q-functions of the same state under different actions $k_1$ and $k_2$:
\begin{align}\label{Equ:structure_of_Q_functions}
    \frac{1}{\theta} \leq \frac{Q^*(s,a_{k_1})}{Q^*(s,a_{k_2})} \leq \theta,
\end{align}
where $\theta > 1$. This parameter incorporates prior knowledge of the $Q$-functions (if any) or can be estimated numerically.

\subsection{Bounds on the coverage coefficient}

Please see \cite{proof} for the complete proofs of all propositions.

\begin{proposition} \normalfont \label{prop: bounds_on_nth_env}
Let $\pi=\pi^{(n)}$ (estimated policy of the $n^{th}$ environment in nEQL algorithm) in (\ref{Equ:CC}). Then:
\begin{align}
    \mathbb{E}[\log C^{\pi^{(n)}}(s,a)] &\leq \log(1\text{+}\theta) + \frac{\lambda^2_n}{3Q^{*}(s,a)^2}\big[\frac{1}{2} - \frac{1}{(1\text{+}\theta)^2}\big]. \nonumber\\
    \mathbb{V}[\log C^{\pi^{(n)}}(s,a)] &\leq \frac{\lambda_n^2}{3Q^{*}(s,a)^2}\Big[1 + \frac{2\theta^2}{(1\text{+}\theta)^2} + \frac{2\sqrt{2}\theta}{1\text{+}\theta} \Big]
\end{align}
\end{proposition}

These bounds are governed by the estimation error variance $\lambda^2_n$, which varies across environments and will be key to optimizing the nEQL algorithm. As $\lambda^2_n$ decreases, the expectation bound tightens, and the variance bound approaches 0. The parameter $\theta$ provides a trade-off between the tightness of the bounds and the flexibility of the assumptions on the Q-functions (\ref{Equ:structure_of_Q_functions}). For example, the constant term $\log(1+\theta)$ can be made smaller, and bounds can be further tightened by employing a more strict assumption than (\ref{Equ:structure_of_Q_functions}) or employing a different exploration strategy than (\ref{Equ: linear_action_selection}) (see \cite{proof} for an example).

\begin{proposition} \normalfont \label{prop: bounds_on_ensemble}
Let $\pi=\hat{\pi}$ (estimated policy of the nEQL algorithm) in (\ref{Equ:CC}). Under the assumptions of Section \ref{Subsec: assumptions}:
\begin{align}
    \mathbb{E}[\log C^{\hat{\pi}}(s,a)] &\leq \log(1\text{+}\theta) \text{+} \frac{\lambda^2}{3Q^{*}(s,a)^2}\Big[\frac{1}{2} \text{-} \frac{1}{(1\text{+}\theta)^2}\Big]\frac{1\text{-}u}{1\text{+}u}.\nonumber\\
    \mathbb{V}[\log C^{\hat{\pi}}(s,a)] &\leq \frac{\lambda^2}{3Q^{*}(s,a)^2}\Big[1 \text{+} \frac{2\theta^2}{(1\text{+}\theta)^2} \text{+} \frac{2\sqrt{2}\theta}{1\text{+}\theta} \Big]\frac{1\text{-}u}{1\text{+}u}.\label{Equ:variance_bound_it}
\end{align}
where $u \in (0,1)$ is the update ratio of nEQL and $\lambda = \max_n \lambda_n$ (The parameters of the nEQL algorithm are outlined in \cite{pn_journal}). These bounds, akin to those in Proposition \ref{prop: bounds_on_nth_env}, depend on the hyper-parameter $u$, allowing for optimization through fine-tuning. Following a strategy as in \cite{pn_journal} (choosing a large $u$ or time-dependent $u_t$ such that $u\xrightarrow[]{t \rightarrow \infty}1$) tightens both bounds. The environment with the largest estimation error variance (\textit{the worst environment}) governs both bounds; thus, optimizing only the Q-learning of the worst environment is an efficient way of improving the performance of the nEQL algorithm.
\end{proposition}

\begin{proposition} \normalfont \label{prop: bounds_on_K}
The bounds in Proposition \ref{prop: bounds_on_ensemble} can be refined to depend on $K$ as follows:
\begin{align}
    \mathbb{E}[\log C^{\hat{\pi}}(s,a)] &\leq \log(1\text{+}\theta) \text{+} \frac{1}{Q^{*}(s,a)^2}\Big[\frac{1}{2} \text{-} \frac{1}{(1\text{+}\theta)^2}\Big]\frac{f(\lambda, u)}{K}.\nonumber\\
    \mathbb{V}[\log C^{\hat{\pi}}(s,a)] &\hspace{-2pt}\leq\hspace{-2pt} \frac{1}{Q^{*}(s,a)^2}\Big[1 \text{+} \frac{2\theta^2}{(1\text{+}\theta)^2} \text{+} \frac{2\sqrt{2}\theta}{1\text{+}\theta} \Big]\frac{f(\lambda, u)}{K}\label{Equ:variance_bound_K}
\end{align}
where $f$ is some function of $\lambda$ and $u$. As $K$ increases, the expectation bound tightens, and the variance bound approaches 0, which aligns with the variance-reduction goal of traditional ensemble algorithms. These bounds can also be optimized by fine-tuning $u$, function $f$, or parameter $\theta$.
\end{proposition}

\subsection{Comparing the bounds of different environments}

The bounds presented in the previous section all rely on the distributional parameters $\lambda_n$ (and $\lambda$). Thus, understanding the relationship between these parameters can be used to improve the accuracy and complexity of nEQL. Proposition 4 of \cite{pn_journal} provides a partial ordering that helps compare estimation error variances in certain environments such as $\lambda_1 \geq \lambda_2 \geq \lambda_4$, $\lambda_1 \geq \lambda_3$ or $\lambda_1 \geq \lambda_5$. However, it provides no information regarding $\lambda_2$ vs $\lambda_3$ or $\lambda_3$ vs $\lambda_5$. Hence, identifying the optimal set of environments may require exhaustive search, which can be computationally expensive for large $K$. Herein, we present several results and an algorithm to compare estimation error variances between any two environments that will be used to determine the optimal set of environments in the nEQL algorithm.

\begin{proposition} \normalfont \label{prop: lambda_1_minimum}
The previous bounds on the expectation and variance are tightest for $n = 1$ and $\lambda_1 = \min_n \lambda_n$. Additionally, $\lambda_n$ is non-monotonic across $n$ for $n > 1$.
\end{proposition}

This proposition suggests that the original environment, with the smallest estimation error variance (\textit{the best environment}), should always be included in the nEQL algorithm as it is most likely to achieve CC close to 1. For $n > 1$, the non-monotonicity of $\lambda_n$ poses challenges in optimizing nEQL, necessitating a strategy to determine the relationship between the $\lambda_n$. This result aligns with Proposition 4 of \cite{pn_journal}.

\begin{proposition} \normalfont \label{prop: algorithm_lambda_n}
Let $\lambda_n$ and $\lambda_m$ be the estimation error variance of two distinct environments. Let the cost function in the underlying MDP be bounded as $c_a(s) \in [c_{min}, c_{max}]$ with $c_{min} > 0$ and $c_{max} < \infty$. Then, the following rule can be employed to compare $\lambda_n$ and $\lambda_m$:\vspace{-5pt}
\begin{align}
&\lambda_n < \lambda_{m} \quad \text{if} \quad f(\gamma, n, m) > \alpha\frac{c_{\text{max}}}{c_{\text{min}}} + (1-\alpha)\frac{c_{\text{min}}}{c_{\text{max}}}. \nonumber\\
&\lambda_n > \lambda_{m} \quad \text{otherwise},
\end{align}
where $f(\gamma, n, m) = \frac{(1-\gamma^{n})(1-\gamma^{m-1})}{(1-\gamma^{m})(1-\gamma^{n-1})}$, $\gamma$ is the common discount factor and $\alpha \in (0,1)$ is the weight factor. This proposition gives complete ordering between all environments (as opposed to the partial ordering of \cite{pn_journal}). This strategy needs at most ${K}\choose{2}$ comparisons to order all $K$ environments. Herein, $\alpha$ is used to determine the decision boundary (see \cite{proof} for details). It can be randomly chosen between (0,1) or can be fine-tuned to minimize the misordering error. 
\end{proposition}

After establishing the bounds on CC and a strategy to order the estimation error variances of different environments, we give the main algorithm of this work in Algorithm \ref{Algo: cc_based_q_learning}. The inputs are the MDP model with $\mathcal{S}, \mathcal{A}$ and estimated transition probabilities ($\hat{p}$) and cost functions ($\hat{c}$) (see \cite{pn_journal} for estimation details), the number of environments ($K$) to initialize the nEQL algorithm out of $K_{total}$ environments, the update ratio of nEQL ($u$), the discount factor ($\gamma$) and a weight factor ($\alpha$). The outputs are the ensemble Q-function output of nEQL ($\hat{Q}$) and corresponding estimated policy $\hat{\bm{\pi}}$. Please refer to \cite{pn_journal} for the details of nEQL. The parameter $K_{total}$ is chosen moderately large and increases as a function of the network size as explained in \cite{pn_journal} and $K \ll K_{total}$. While exact bounds in Propositions \ref{prop: bounds_on_nth_env}, \ref{prop: bounds_on_ensemble}, \ref{prop: bounds_on_K} require the optimal Q-functions ($Q^*$), we can order the environments without knowing optimal Q-functions as $Q^*$ is a constant with respect to $n$.

\algdef{SE}[REPEATN]{RepeatN}{End}[1]{\algorithmicrepeat\ #1 \textbf{times}}{\algorithmicend}
\begin{algorithm}[t]
\caption{Coverage-based ensemble Q-learning (CCQ)}
\hspace*{\algorithmicindent} \textbf{Inputs:} MDP model ($\mathcal{S}, \mathcal{A}$, $\hat{p}$, $\hat{c}$), $K$, $K_{total}$, $\gamma$, $u$, $\alpha$ \\
\hspace*{\algorithmicindent} \textbf{Outputs:} $\hat{Q}$, $\hat{\pi}$ 
\begin{algorithmic}[1]
    \State Estimate $c_{min}$, $c_{max}$ using estimated cost functions ($\hat{c}$)
    \State Use Proposition 5 to sort $\lambda_n$ for $n = 1,2,...,K_{total}$ in increasing order.
    \State Pick the first $K$ environments with the tightest bounds on the coverage coefficient using Proposition 1 ($K$ $\ll$ $K_{total}$)
    \State Run nEQL \cite{pn_journal} with the $K$ environments (the other hyper-parameters are optimized as in  \cite{pn_journal}).
    \State Output estimated Q-functions $\hat{Q}$ and estimated policy $\hat{\pi}$
\end{algorithmic}
\label{Algo: cc_based_q_learning}
\end{algorithm}

\section{Numerical Results}
\label{sec:numerical_results}
\begin{figure}[t]
    \centering
    \subfloat[\footnotesize MISO energy harvesting wireless network with multiple relays]{{\includegraphics[width=4.2cm]{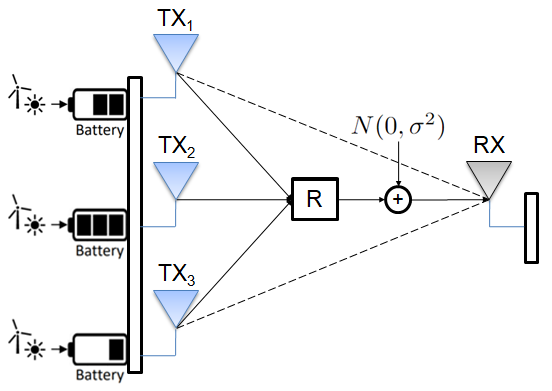}}}%
    \hspace{2pt}
    \subfloat[\footnotesize MIMO wireless network]{{\includegraphics[width=4cm]{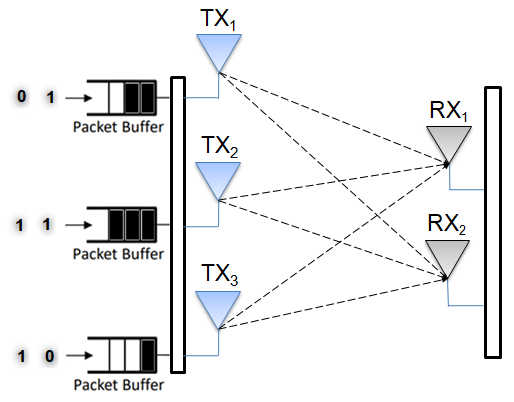}}}%
    \caption{\footnotesize Examples wireless network models.}
    \label{fig:networks}
\end{figure}

We consider two distinct wireless network models that vary in their topology, scale, objective, and implementations: (i) MISO energy harvesting wireless network with multiple relays \cite{colink_journal}, (ii) MIMO wireless network \cite{ln_journal}. Example wireless networks are shown in Fig. \ref{fig:networks}. We use the following numerical parameters: $|\mathcal{S}| = 10000$, $|\mathcal{A}| = 2$, $K = 5$, $K_{total} = 10$, $u = 0.5$, $\gamma = 0.95$, $\alpha \sim \operatorname{unif}(0,1)$, $(s,a) = (6,1)$. The parameter $\theta$ is numerically estimated and ranges in $[1.04, 1.26]$ for different settings. In the wireless networks, the $n^{th}$ environments consider $n$ packet changes at a time -- transmitting or receiving $n$ energy/data packets, and the corresponding PTTs describe the $n$-step transition probabilities.

For MISO network, Algorithm 1 gives the following set in increasing order: $\{\lambda_1, \lambda_2, \lambda_3, \lambda_6, \lambda_5, \lambda_4, \lambda_7, \lambda_9, \lambda_8, \lambda_{10}\}$. In contrast, the partial ordering approach of \cite{pn_journal} gives us the following three set of orders: $\{\lambda_1, \lambda_2, \lambda_4, \lambda_8\}$, $\{\lambda_1, \lambda_3, \lambda_6\}$ and $\{\lambda_1, \lambda_5, \lambda_{10}\}$ each with no preference. Extensive simulations yield the following true ordering: $\{\lambda_1, \lambda_2, \lambda_3, \lambda_5, \lambda_6, \lambda_4, \lambda_7, \lambda_9, \lambda_8, \lambda_{10}\}$. Observe that Algorithm 1 gives the correct set of environments for $K=5$ $(1,2,3,5,6)$, yet the partial ordering is mostly inconclusive. We repeat the simulation using the following parameters: $|\mathcal{S}|$ $\in$ $\{500,1000,...,40000\}$, $\mathcal{A} \in$ [2,6], $K_{total} \in$ [10,20], $K \in$ [2,10] using both MISO and MIMO networks, the number of transmitters, receivers and relays $\in [2,8]$. Overall, Algorithm 1 gives us 82\% accuracy in finding the set of optimal environments, which shows the robustness of the algorithm to the changes in network parameters. Algorithm 1 is 45\% faster than the nEQL algorithm with partial ordering \cite{pn_journal} (with exhaustive search where the partial ordering is inconclusive).
\addtolength{\topmargin}{+0.15cm}

\begin{figure}[t]
    \centering
    \subfloat[\footnotesize APE of different algorithms across increasing state-space size \label{fig:APE}]{{\includegraphics[width=4.1cm]{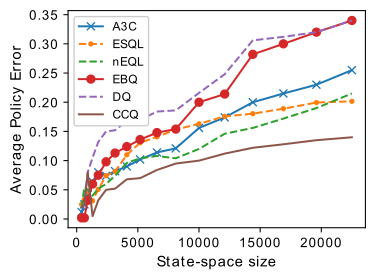}}}
    \hspace{2pt}
    \subfloat[\footnotesize $C^{\pi^{(n)}}(6,1)$ vs environment order $n$ across iterations \label{fig:cc_with_n} ]{{\includegraphics[width=3.9cm]{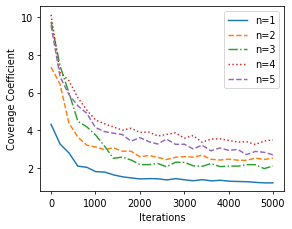}}}
    \vspace{-2pt}
    \subfloat[\footnotesize $\log C^{\pi^{(1)}}(6,1)$ and upper bound from Proposition 1\label{fig:cc_with_n_for_1} ]{{\includegraphics[width=3.9cm]{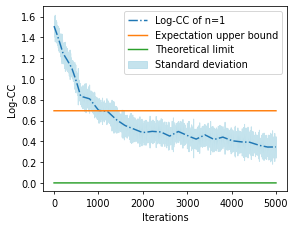}}}
    \hspace{2pt}
    \subfloat[\footnotesize $\log C^{\hat{\pi}}(4,0)$ and upper bound from Proposition 2 \label{fig:cc_expectation}]{{\includegraphics[width=3.9cm]{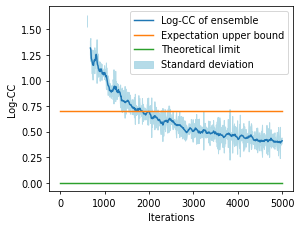}}}
    \caption{\footnotesize Numerical simulations}
\end{figure}

We also evaluate the performance of Algorithm 1 using MIMO network using Average Policy Error (APE), defined as $\frac{1}{|\mathcal{S}|} \sum_{s=1}^{|\mathcal{S}|} \mathbf{1}\left(\bm{\pi^{*}}(s) \neq \hat{\bm{\pi}}(s)\right)$, where $\pi^*$ is the optimal policy of the original environment as defined in Section \ref{subsec:q_learning}. For comparison, we employ the following Q-learning algorithms: (i) n-hop Ensemble Q-Learning (nEQL) \cite{pn_journal}, (ii) Ensemble Synthetic Q-Learning (ESQL) \cite{ln_journal}, (iii) Asynchronous Advantage Actor Critic (A3C) \cite{mnih2016asynchronous}, (iv) Ensemble Bootstrapping Q (EBQ) \cite{ensemble_bootstrap_q}, and (v) Double Q (DQ) \cite{double_q}. These algorithms differ in their objective, estimation strategy, and data collection strategy and provide a fair comparison. The APE results vs state-space size is shown in Fig.\ref{fig:APE}. The proposed algorithm can reduce the APE of nEQL by \%50 and also achieves \%40 less APE than the other algorithms across large state-spaces, which shows the benefits of initializing the algorithm with the optimal set of environments.

We give the CC over time across different orders (n = 1,2,3,4,5) for $(s,a)=(6,1)$ in Fig.\ref{fig:cc_with_n} using the MIMO network with $\theta = 1.15$. The CC varies non-monotonically across n, with the original environment having the smallest CC, followed by the 3rd, 2nd, 5th, and 4th environments. The estimation error variances of the environments from Proposition 5 and the numerically estimated variances follow the same order as the CCs ($\lambda_1 \leq \lambda_3 \leq \lambda_2 \leq \lambda_5 \leq \lambda_4$). This suggests that estimation error variances can effectively order CCs. The log-CC under the estimated policy of the original environment ($\pi^{(1)}$) for $(6,1)$, along with expectation bound from Proposition 1, is shown in Fig. \ref{fig:cc_with_n_for_1}. The upper bound is useful when the algorithm converges, and its tightness changes with $(s,a)$ and $n$. Fig. \ref{fig:cc_expectation} illustrates similar trends for $(s,a) = (4,0)$ under the policy of Algorithm 1 ($\hat{\pi}$) with the expectation bound from Proposition 2.

\vspace{-6pt}
\section{Conclusions}\label{sec:conclusion}
\vspace{-4pt}
We provided a detailed analysis of the data coverage conditions of recently proposed multi-environment ensemble Q-learning algorithms \cite{pn_journal, ln_journal} for optimizing a variety of real-world wireless networks. We provided upper bounds on the expectation and variance of different coverage coefficients and explain how to interpret them. Exploiting these bounds, we presented an algorithm to provide an efficient and accurate way of initializing these algorithms by ordering the utilities of different environments. Our simulations on two different wireless networks showed that we could reduce the policy error by \%50 and increase the speed by \%40 of the state-of-the-art prior work. We also verified our theoretical assumptions.

\bibliographystyle{unsrt}
\bibliography{references.bib}

\begin{thebibliography}{10}

\bibitem{talha_jie_asilomar}
Jie Wang, Talha Bozkus, Yao Xie, and Urbashi Mitra.
\newblock Reliable adaptive recoding for batched network coding with burst-noise channels.
\newblock In {\em 2023 57th Asilomar Conference on Signals, Systems, and Computers}, pages 220--224. IEEE, 2023.

\bibitem{colink_journal}
Talha Bozkus and Urbashi Mitra.
\newblock Link analysis for solving multiple-access mdps with large state spaces.
\newblock {\em IEEE Transactions on Signal Processing}, 71:947--962, 2023.

\bibitem{ln_journal}
Talha Bozkus and Urbashi Mitra.
\newblock Leveraging digital cousins for ensemble q-learning in large-scale wireless networks.
\newblock {\em IEEE Transactions on Signal Processing}, 72:1114--1129, 2024.

\bibitem{talha_asilomar}
Talha Bozkus and Urbashi Mitra.
\newblock A novel ensemble q-learning algorithm for policy optimization in large-scale networks.
\newblock In {\em 2023 57th Asilomar Conference on Signals, Systems, and Computers}, pages 1381--1386. IEEE, 2023.

\bibitem{bertsekas_book}
Dimitri Bertsekas.
\newblock {\em Reinforcement learning and optimal control}.
\newblock Athena Scientific, 2019.

\bibitem{speedy_q}
M.~G. Azar, R.~Munos, M.~Ghavamzadaeh, and H.~J Kappen.
\newblock Speedy q-learning.
\newblock 2011.

\bibitem{fitted_q}
Martin Riedmiller.
\newblock Neural fitted q iteration--first experiences with a data efficient neural reinforcement learning method.
\newblock In {\em European conference on machine learning}, pages 317--328. Springer, 2005.

\bibitem{double_q}
Hado Hasselt.
\newblock Double q-learning.
\newblock {\em Advances in neural information processing systems}, 23, 2010.

\bibitem{averaged_dqn}
Oron Anschel, Nir Baram, and Nahum Shimkin.
\newblock Averaged-dqn: Variance reduction and stabilization for deep reinforcement learning.
\newblock In {\em International conference on machine learning}, pages 176--185. PMLR, 2017.

\bibitem{ensemble_bootstrap_q}
Oren Peer, Chen Tessler, Nadav Merlis, and Ron Meir.
\newblock Ensemble bootstrapping for q-learning.
\newblock In {\em International Conference on Machine Learning}, pages 8454--8463. PMLR, 2021.

\bibitem{randomized_double_q}
Xinyue Chen, Che Wang, Zijian Zhou, and Keith~W. Ross.
\newblock Randomized ensembled double q-learning: Learning fast without a model.
\newblock {\em CoRR}, abs/2101.05982, 2021.

\bibitem{pn_journal}
Talha Bozkus and Urbashi Mitra.
\newblock Multi-timescale ensemble $q$-learning for markov decision process policy optimization.
\newblock {\em IEEE Transactions on Signal Processing}, 72:1427--1442, 2024.

\bibitem{online_q_learning_1}
Chi Jin, Zeyuan Allen-Zhu, Sebastien Bubeck, and Michael~I Jordan.
\newblock Is q-learning provably efficient?
\newblock {\em Advances in neural information processing systems}, 31, 2018.

\bibitem{online_q_learning_2}
Yue Wang and Shaofeng Zou.
\newblock Online robust reinforcement learning with model uncertainty.
\newblock {\em Advances in Neural Information Processing Systems}, 34:7193--7206, 2021.

\bibitem{offline_q_learning_1}
Aviral Kumar, Aurick Zhou, George Tucker, and Sergey Levine.
\newblock Conservative q-learning for offline reinforcement learning.
\newblock {\em Advances in Neural Information Processing Systems}, 33:1179--1191, 2020.

\bibitem{offline_q_learning_2}
Rishabh Agarwal, Dale Schuurmans, and Mohammad Norouzi.
\newblock An optimistic perspective on offline reinforcement learning.
\newblock In {\em International Conference on Machine Learning}, pages 104--114. PMLR, 2020.

\bibitem{hybrid_q_1}
Yuda Song, Yifei Zhou, Ayush Sekhari, J~Andrew Bagnell, Akshay Krishnamurthy, and Wen Sun.
\newblock Hybrid rl: Using both offline and online data can make rl efficient.
\newblock {\em arXiv preprint arXiv:2210.06718}, 2022.

\bibitem{hybrid_q_2}
Todd Hester, Matej Vecerik, Olivier Pietquin, Marc Lanctot, Tom Schaul, Bilal Piot, Dan Horgan, John Quan, Andrew Sendonaris, Ian Osband, et~al.
\newblock Deep q-learning from demonstrations.
\newblock In {\em Proceedings of the AAAI conference on artificial intelligence}, volume~32, 2018.

\bibitem{coverage_1}
Tengyang Xie, Dylan~J Foster, Yu~Bai, Nan Jiang, and Sham~M Kakade.
\newblock The role of coverage in online reinforcement learning.
\newblock {\em arXiv preprint arXiv:2210.04157}, 2022.

\bibitem{coverage_2}
Masatoshi Uehara and Wen Sun.
\newblock Pessimistic model-based offline reinforcement learning under partial coverage.
\newblock {\em arXiv preprint arXiv:2107.06226}, 2021.

\bibitem{q_learning_convergence}
Francisco~S Melo.
\newblock Convergence of q-learning: A simple proof.
\newblock {\em Institute Of Systems and Robotics, Tech. Rep}, pages 1--4, 2001.

\bibitem{coverage_3}
Fanghui Liu, Luca Viano, and Volkan Cevher.
\newblock What can online reinforcement learning with function approximation benefit from general coverage conditions?
\newblock In {\em International Conference on Machine Learning}, pages 22063--22091. PMLR, 2023.

\bibitem{uniform_assump_1}
Sebastian Thrun and Anton Schwartz.
\newblock Issues in using function approximation for reinforcement learning.
\newblock In {\em Proceedings of the 1993 Connectionist Models Summer School Hillsdale, NJ. Lawrence Erlbaum}, volume~6, pages 1--9, 1993.

\bibitem{proof}
Talha Bozkus and Urbashi Mitra.
\newblock Supplementary appendix file.
\newblock \url{https://github.com/talhabozkus/spawc-24-supplementary-material}, 2024.

\bibitem{mnih2016asynchronous}
Volodymyr Mnih, Adria~Puigdomenech Badia, Mehdi Mirza, Alex Graves, Timothy Lillicrap, Tim Harley, David Silver, and Koray Kavukcuoglu.
\newblock Asynchronous methods for deep reinforcement learning.
\newblock In {\em International conference on machine learning}, pages 1928--1937. PMLR, 2016.

\end{thebibliography}

\end{document}